\documentclass[prd,aps,,preprintnumbers,floats,floatfix,nofootinbib]{revtex4}
\usepackage{graphicx,amsmath}
\usepackage{amssymb}
\usepackage{subfigure}

\begin{document}

\title{THE TOP TRIANGLE MOOSE\footnote{This manuscript will appear in the proceedings of SCGT09; the speaker at the conference was E.H. Simmons}}

\author{R. S. CHIVUKULA, N. D. CHRISTENSEN, B. COLEPPA and E. H. SIMMONS$^*$}

\address{Department of Physics and Astronomy, Michigan State University\\
East Lansing, Michigan, 48824, USA\\
$^*$E-mail: esimmons@msu.edu\\
http://www.pa.msu.edu/hep/hept/index.php}

\begin{abstract}
We introduce a deconstructed model that incorporates both Higgsless and top-color mechanisms. The model alleviates the typical tension in Higgsless models between obtaining the correct top quark mass and keeping $\Delta\rho$ small. It does so by singling out the top quark mass generation as arising from a Yukawa coupling to an effective top-Higgs which develops a small vacuum expectation value, while electroweak symmetry breaking results largely from a Higgsless mechanism. As a result, the heavy partners of the SM fermions can be light enough
to be seen at the LHC. 
\end{abstract}


\maketitle

\section{Introduction}
 Higgsless models \cite{Csaki-Reference} have recently emerged as a novel way of understanding the mechanism of electroweak symmetry breaking (EWSB) without the presence of a scalar particle in the spectrum. In an extra dimensional context, these can be understood in terms of a $SU(2)\times SU(2)\times U(1)$ gauge theory in the bulk of a finite $AdS$ spacetime \cite{Csaki-Higgsless,Csaki-Higglsess2,Nomura-Higgsless,Sundrum-Higgsless}, with symmetry breaking encoded in the boundary conditions of the gauge fields.  One can understand the low energy properties of such theories in a purely four dimensional picture by invoking the idea of deconstruction
\cite{Deconstruction-Georgi,Deconstruction-Hill}. The ``bulk'' of the extra dimension is replaced by a chain of gauge groups strung together by non linear sigma model fields. The spectrum typically includes extra sets of charged and neutral vector bosons and heavy fermions. A general analysis of Higgsless models \cite{Delocalization-1,Delocalization-2,Delocalization-3,Delocalization-4,Casalbuoni:2005rs,Delocalization-5} suggests that to satisfy precision electroweak constraints, the standard model (SM) fermions must be `delocalized' into the bulk.  A useful realization of this idea, ``ideal fermion delocalization'" \cite{IDF}, dictates that the light fermions be delocalized so as not to couple to the heavy charged gauge bosons. The simplest framework capturing these ideas is the ``three site Higgsless model''\cite{three site ref}, with just one gauge group in the bulk and correspondingly, only one set of heavy vector bosons. The twin constraints of getting the correct value of the top quark mass and having an admissible $\rho$ parameter push the heavy fermion masses into the TeV regime \cite{three site ref} in that model.

This presentation summarizes Ref.~\cite{Chivukula:2009ck}, in which we seek to decouple these constraints by separating the mechanisms that break the electroweak symmetry and generate the masses of the third family of fermions. In this way, one can obtain a massive top quark and heavy fermions in the sub TeV region, without altering tree level electroweak predictions. To present a minimal model with these features, we modify the three site model by adding a ``top Higgs'' field, $\Phi,$ that couples preferentially to the top quark.  The resulting model is shown in Moose notation \cite{Moose} in Figure 1; we will refer to it as the ``top triangle moose.'' 

The idea of a top Higgs is motivated by top condensation models (see references in Ref.~\cite{Chivukula:2009ck}), and the specific framework shown here is most closely aligned with topcolor assisted technicolor theories first proposed in Ref.~\cite{Hill-TC2},  in which EWSB occurs via technicolor\cite{Eichten:1979ah, Dimopoulos:1979es} interactions while the top mass has a dynamical component arising from topcolor \cite{Hill-Topcolor-1,Hill-Topcolor-2} interactions and a small component generated by an extended technicolor mechanism.   The dynamical bound state arising from topcolor dynamics can be identified as a composite top Higgs field, and the low-energy spectrum includes a top Higgs boson. The extra link in our triangle moose that corresponds to the top Higgs field results in the presence of uneaten Goldstone bosons, the top pions, which couple preferentially to the third generation. The model can thus be thought of as the deconstructed version of a topcolor assisted technicolor model.

\section{The Model}

We now introduce the essential features of the model, which are required in order to understand the LHC phenomenology.  Full details are presented in Ref.~\cite{Chivukula:2009ck}.

\begin{figure}[t]
\begin{center}
\includegraphics[width=1.25in]{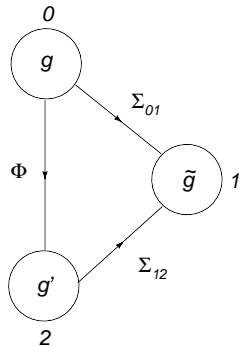}
\caption{The $SU(2) \times SU(2) \times U(1)$ gauge structure of the model in Moose notation \cite{Moose}. The $SU(2)$ coupling $g$ and $U(1)$ coupling $g'$ of sites 0 and 2 are approximately the SM $SU(2)$ and hypercharge gauge couplings, while the $SU(2)$ coupling 
$\tilde{g}$ represents the 'bulk' gauge coupling. 
}
\label{fig:Triangle}
\end{center}
\end{figure}

The electroweak gauge structure of our model is $SU(2)_0\times SU(2)_1\times U(1)_2$.  This is shown using Moose notation \cite{Moose} in Figure \ref{fig:Triangle}, in which the $SU(2)$ groups are associated with sites 0 and 1, and the $U(1)$ group is associated with site 2. The SM fermions deriving their $SU(2)$ charges mostly from site 0 (which is most closely associated with the SM $SU(2)$) and the bulk fermions mostly from site 1. The extended electroweak gauge structure of the theory is the same as that of the BESS models \cite{BESS-1,BESS-2}, motivated by models of hidden local symmetry \cite{HLS-1,HLS-2,HLS-3,HLS-4,HLS-5}.

The non linear sigma field $\Sigma_{01}$ breaks the $SU(2)_0\times SU(2)_1$ gauge symmetry down to $SU(2)$, and field $\Sigma_{12}$ breaks $SU(2)_1 \times U(1)_2$ down to $U(1)$. The left handed fermions are $SU(2)$ doublets residing at sites 0 ($\psi_{L0}$) and 1 ($\psi_{L1}$), while the right handed fermions are a doublet under $SU(2)_{1}$($\psi_{R1}$) and two $SU(2)$-singlet fermions at site 2 ($u_{R2}$ and $d_{R2}$). The fermions $\psi_{L0}$, $\psi_{L1}$, and $\psi_{R1}$ have SM-like $U(1)$ charges ($Y$):  $+1/6$ for quarks and $-1/2$ for leptons. Similarly, the fermion $u_{R2}$ ($d_{R2}$) has an SM-like $U(1)$ charge of $+2/3$ ($-1/3$); the right-handed leptons, likewise, have  $U(1)$ charges corresponding to their SM hypercharge values. The third component of isospin, $T_3$, takes values $+1/2$ for ``up'' type fermions and $-1/2$ for ``down'' type fermions, just like in the SM. The electric charges satisfy $Q=T_{3}+Y$.

We add a `top-Higgs' link to separate top quark mass generation from EWSB. The top quark couple preferentially to the top Higgs link via the Largangian:
\begin{equation}
\mathcal{L}_{top}=-\lambda_{t}\bar{\psi}_{L0}\,\Phi\, t_{R}+h.c.\label{top quark mass L}
\end{equation}
When the field $\Phi$ develops a non zero vacuum expectation value, Eqn.(\ref{top quark mass L}) generates a top quark mass term. Since we want most EWSB to come from the Higgsless side, we choose the vacuum expectation values of $\Sigma_{01}$ and $\Sigma_{02}$ to be $F=\sqrt{2}\,v$ $\textrm{cos}\,\omega$  and the one associated with the top Higgs sector to be $f=\langle\Phi\rangle=v$ $\textrm{sin\,}\omega$ (where $\omega$ is small). The top Higgs sector also includes the uneaten Goldstone bosons, the top pions; we assume they
are heavy enough not to affect electroweak phenomenology. 

The mass terms for the light fermions arise from Yukawa couplings of the fermionic fields with the non linear sigma fields
\begin{eqnarray}
\mathcal{L} & = & M_{D}\left[\epsilon_{L}\bar{\psi}_{L0}\Sigma_{01}\psi_{R1}+\bar{\psi}_{R1}\psi_{L1}+\bar{\psi}_{L1}\Sigma_{12}\left(\begin{array}{cc}
\epsilon_{uR} & 0\\
0 & \epsilon_{dR}\end{array}\right)\left(\begin{array}{c}
u_{R2}\\
d_{R2}\end{array}\right)\right].
\label{eqn:Light fermion mass}
\end{eqnarray}
We denote the Dirac mass  setting the scale of the heavy fermion masses as $M_D$.  Here, $\epsilon_{L}$ is a flavor-universal parameter describing delocalization of the left handed fermions. All the flavor violation for the light fermions is encoded in the last term; the delocalization parameters for the right handed fermions, $\epsilon_{fR}$, can be adjusted to realize the masses and mixings of the up and down type fermions. For our phenomenological study, we will, for the most part, assume that all the fermions, except the top, are massless and hence will set these $\epsilon_{fR}$ parameters to zero.  

The \emph{tree level} contributions to precision measurements in Higgsless models come from the coupling of standard model fermions to the heavy gauge bosons.  Choosing the profile of a light fermion bilinear along the Moose to be proportional to the profile of the light $W$ boson makes the fermion current's coupling to the $W'$ vanish because the $W$ and $W'$ fields are mutually orthogonal. This procedure (called ideal fermion delocalization \cite{IDF})  keeps deviations from the SM values of all electroweak quantities at a phenomenologically acceptable level.  We find that the ideal delocalization condition in this model is $\epsilon_{L}^{2}=M_W^2 / 2 M_{W'}^2$, as in the three-site model.

The top quark mass matrix may be read from Eqns. (\ref{top quark mass L}) and (\ref{eqn:Light fermion mass}) and is given by:
\begin{equation}
\left(\begin{array}{cc}
M_{D}\epsilon_{tL} & \lambda_{t}v\textrm{sin}\omega\\
M_{D} & M_{D}\epsilon_{tR}\end{array}\right).
\label{top mass matrix}
\end{equation}
Diagonalizing the top quark mass matrix perturbatively in $\epsilon_{tL}$
and $\epsilon_{tR}$, we find the mass of the top quark is:
\begin{equation}
m_{t}= \lambda_{t}v\,\textrm{sin}\,\omega\left[1+\frac{\epsilon_{tL}^{2}+\epsilon_{tR}^{2}+\frac{2}{a}\epsilon_{tL}\epsilon_{tR}}{2(-1+a^{2})}\right], \qquad\qquad a\equiv\frac{\lambda_{t}\, v\,\textrm{sin}\omega}{M_{D}},
\label{top mass}
\end{equation}
Thus, we see that $m_{t}$ depends mainly on $v$ and only slightly on $\epsilon_{tR}$, in contrast to the situation in the three-site model where $m_t \propto M_D \epsilon_L \epsilon_{tR}$. 

Since the $b_L$ is the $SU(2)$ partner of the $t_L$, its delocalization is (to the extent that
$\epsilon_{bR}\simeq 0$) also determined by $\epsilon_{tL}$. Thus, the tree
level value of the $Zb\bar{_{L}b_{L}}$coupling can be used to constrain
$\epsilon_{tL}$. We find $g_{L}^{Zbb}$ equals its tree-level SM value if the left-handed top quark is  delocalized exactly as the light fermions are: $\epsilon_{tL} = \epsilon_L$.

Finally, the contribution of the heavy top-bottom doublet to $\Delta\rho$ is of the same form as in the three-site model \cite{three site ref}: $\Delta\rho= M_{D}^{2}\,\epsilon_{tR}^{4} / 16\,\pi^{2}\, v^{2}$.
The key difference is that, since the top quark mass is dominated by the vev of the top Higgs instead of $M_{D}$, $\epsilon_{tR}$ can be as small as the $\epsilon_R$ of any light fermion. There is no conflict between the twin goals of a large top quark mass and a small  value of $\Delta\rho$. Thus, the heavy fermions in the top triangle moose can be light enough to be seen at the LHC.

\section{Heavy quarks at the LHC}

We now summarize our analysis\cite{Chivukula:2009ck} of the possible discovery modes of the heavy quarks at the LHC; this work employed the CalcHEP package \cite{Pukhov-Calchep}. 

\subsection{Pair production: $pp\rightarrow Q\bar{Q}\rightarrow WZqq\rightarrow lll\nu jj$} \label{Subsection:pair}

Pair production of heavy quarks occurs at LHC via gluon fusion and quark annihilation processes, with the former dominating for smaller $M_D$.  Each heavy quark decays to a vector boson and a light fermion. For $M_{D}<M_{W',Z'}$, the decay is purely to the standard model gauge bosons. We study the case where one heavy quark decays to $Z+j$ and the other decays to $W+j$, with the gauge bosons subsequently decaying leptonically. Thus, the final state is $lll \nu jj$.

To enhance the signal to background ratio, we have imposed a variety of cuts, as shown in Table 1. We note that the the two jets in the signal should have a high $p_{T}\,$($\sim M_{D}/2)$, since they each come from the 2-body decay of a heavy fermion. Thus, imposing strong $p_{T}$ cuts on the outgoing jets can eliminate much of the SM background without affecting the signal too much. We also expect the $\eta$ distribution of the jets to be largely central, which suggests an $\eta$ cut: $|\eta|\leq2.5$. We impose standard separation cuts between the two jets and between jets and leptons to ensure that they are observed as distinct final state particles.   We also  impose basic identification cuts on the leptons and missing transverse energy.

 We identify the leptons that came from the $Z$ by imposing the invariant mass cut $(M_{Z}-2\, \textrm{GeV})<M_{ll}<(M_{Z}+2\,\textrm{GeV})$.  We then combine this lepton pair with a leading-$p_T$ light jet to reconstruct the heavy fermion mass.   Because one cannot know which light jet came from the $Q$, we actually combine the lepton pair first with the light jet of largest $p_T$ and then, separately, with the light jet of next-largest $p_T$, and include both reconstructed versions of each event in our analysis.  This yields\cite{Chivukula:2009ck} an invariant mass distribution with a narrow signal peak standing out cleanly at $M_D$ above a tiny ``background" from the wrongly-reconstructed signal events.

\begin{table}[t,b]
  \caption{Cuts employed in the pair (left) and single (right) production channels for the heavy quarks.
 $\Delta R_{jj}=\sqrt{\Delta\eta_{jj} + \Delta \phi_{jj}}$ refers to the separation between the two jets; $\Delta R_{jl}$ refers to the angular separation between a lepton and a jet.} 
{{ \begin{tabular}{| c | c | }
    \hline
    Variable & Cut \\ \hline\hline
    $p_{Tj}$ & $>$100 \textrm{GeV} \\ \hline
    $p_{Tl}$ & $>$15 \textrm{GeV} \\ \hline
    \textrm{Missing} $E_{T}$ & $>$15 \textrm{GeV} \\ \hline
    $|\eta_{j}|$ & $<$ 2.5 \\ \hline
    $|\eta_{l}|$ & $<$ 2.5 \\ \hline
    $\Delta R_{jj}$ & $>$0.4 \\ \hline
    $\Delta R_{jl}$ & $>$0.4 \\ \hline
    $M_{ll}$ & 89 \textrm{GeV}$<M_{ll}<93$ \textrm{GeV} \\
    \hline
  \end{tabular}}
  \hspace{1cm}
   { \begin{tabular}{| c | c | }
    \hline
    Variable & Cut \\ \hline\hline
   $p_{Tj\ \mbox{hard}}$ & \quad $>$200 \textrm{GeV} \quad \\ \hline
   $p_{Tj\ \mbox{soft}}$ & $>$15 \textrm{GeV} \\ \hline
   $p_{Tl}$ & $>$15 \textrm{GeV} \\ \hline
   Missing $E_{T}$ & $>$15 \textrm{GeV} \\ \hline
   $|\eta_{j\ \mbox{hard}}|$ & $<$ 2.5 \\ \hline
   $|\eta_{j\ \mbox{soft}}|$ & 2$<|\eta|<4$ \\ \hline
   $|\eta_{l}|$ & $<$ 2.5 \\ \hline
   $\Delta R_{jj}$ & $>$0.4 \\ \hline
   $\Delta R_{jl}$ & $>$0.4 \\ \hline
  \end{tabular}}}
\label{tab: cuts pair}
\end{table}

When generating the signal events, we included the four flavors of heavy quarks, $U, D, C, S$, that should have similar phenomenology. We estimate the size of the peak by counting the signal events in the invariant mass window: 
$(M_{D}-10\, \textrm{GeV}) <M_{jll}<  (M_{D}+10\, \textrm{GeV})$.
To analyze the SM background, we fully calculated the irreducible $pp\rightarrow ZWjj$ process and subsequently decayed the $W$ and $Z$ leptonically. Imposing the full set of cuts on the final state $lll \nu jj$ entirely eliminates the SM background. 

We find there are many signal events in the region of parameter space where $Q\rightarrow Vq$ decays are allowed but $Q\rightarrow V'q$ decays are kinematically forbidden. The precise number is controlled by the branching ratio of the heavy fermion into the SM vector bosons. Since the SM background is negligible, if we assume the signal events are Poisson-distributed, then we can take 10 events to represent a 5$\sigma$ signal at 95\% c.l.  Figure \ref{fig:Luminosity plot} shows the results, recast in the form of the integrated luminosity required to achieve a 5$\sigma$ discovery signal.  We see that the pair-production process we have studied spans almost the entire parameter space. However, in the region where $M_{D}\geq \textrm{900 GeV}$ and $M_{W'}\leq M_{D}$ there will not be enough signal events for the discovery of the heavy quark since the decay channel $Q \to W' q$ becomes significant. To explore this region, we now investigate the single production channel where the heavy quark decays to a heavy gauge boson.

\subsection{Single production: $pp\rightarrow Qq\rightarrow W'qq'\rightarrow WZqq'$} \label{Subsection:single}

While the single production channel  is electroweak, the smaller cross section is compensated by the fact that the $u$ and $d$ are valence quarks, and their parton distribution functions fall less sharply than the gluon's.  Also, there is less phase space suppression in the single production channel than in the pair production case. We analyze the processes $\left[u,u\rightarrow u,U \right ]$, $\left [d,d\rightarrow d,D  \right ]$ and $\left [u,d\rightarrow u,D \ {\rm or}\ U,d\right ]$, which occur through a $t$ channel exchange of a $Z$ and $Z'$. Since we want to look at the region of parameter space where $M_{W'}$ is smaller than $M_{D},$ we let the heavy quark decay to a $W'$. The $W'$ decays 100\% of the time to a $W$ and $Z$, because its coupling to two SM fermions is zero in the limit of ideal fermion delocalization. We constrain both the $Z$ and $W$ to decay leptonically so the final state is $lll \nu jj$. 

Again, we expect the jet from the decay of the heavy quark to have a large $p_{T}$, and we impose a strong $p_{T}$ cut on this ``hard jet". As before, this jet should be central, so we impose the same $\eta$ cut on the hard jet.  We expect the $\eta$ distribution of the ``soft jet" arising from the light quark in the production process to be in the forward region, $2<|\eta|<4$. We impose the same $\Delta R$ jet separation and jet-lepton separation cuts as before, along with basic identification cuts on the leptons and missing transverse energy. The complete set of cuts is in the right side of Table 1.

The leptonic $W$ decay introduces the usual two fold ambiguity in determining the neutrino momentum and hence, we have performed a transverse mass analysis of the process, defining the transverse mass variable \cite{Transverse} of interest as:
\begin{equation}
M_{T}^{2}=\left(\sqrt{M^{2}(lllj)+p_{T}^{2}(lllj)}+\left|p_{T}(missing)\right|\right)^{2}-\left|\overrightarrow{p_{T}}(lllj)+\overrightarrow{p_{T}}(missing)\right|^{2}
\label{Transverse mass variable}
\end{equation}
We expect the distribution to fall sharply at $M_{D}$ in the narrow width approximation, and indeed we find that there are typically few or no events beyond $M_{D}+20$ GeV in the distributions. Thus, we take the signal events to be those in the transverse mass window: $(M_{D}-200\, \textrm{GeV})  <M_{T}<  (M_{D}+20\, \textrm{GeV})$.

\begin{figure}[tb]
\begin{center}
\includegraphics[width=3in]{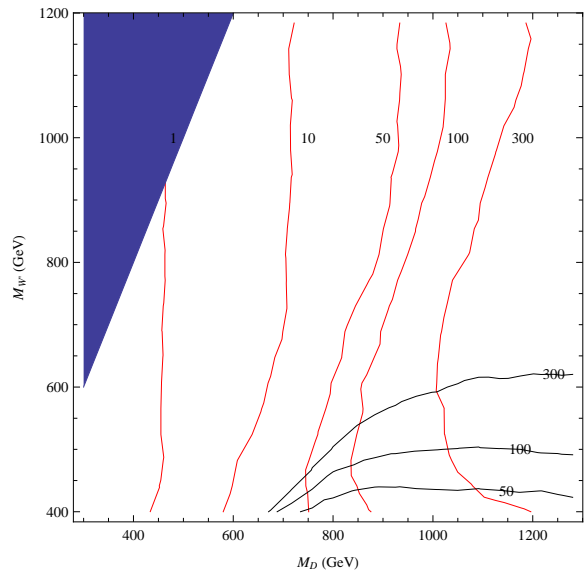}
\caption{Luminosity required for a $5\sigma$ discovery of
the heavy vector fermions at the LHC in the single (blue curves, nearly horizontal) and pair
(red curves, nearly vertical) production channels. The shaded area is non perturbative
and not included in the study. The two channels are
complementary and allow almost the entire region to
be covered in 300 $fb^{-1}.$
}
\label{fig:Luminosity plot}
\end{center}
\end{figure}

The SM background for this process, $pp\rightarrow WZjj \rightarrow jjl\nu ll$, was calculated 
summing over the $u,$ $d,$ $c$, $s$ and gluon jets and the first two
families of leptons. Since we apply a strong $p_{T}$ cut on only one of the jets (unlike in the pair production case), there is a non zero SM background, as plotted in  Ref.~\cite{Chivukula:2009ck}.  The luminosity necessary for a $5\sigma$ discovery at 95\% c.l. can be calculated by requiring $(N_{signal}/\sqrt{N_{bkrnd}}) \geq 5$, as per a Gaussian distribution. Figure \ref{fig:Luminosity plot}) shows the results, again recast in the form of the integrated luminosity required to achieve a 5$\sigma$ discovery signal.  Almost the entire parameter space is covered, with the pair and single production channels nicely complementing each other.

\end{document}